# Effect of Distributed Shield Insertion on Crosstalk in Inductively Coupled VLSI Interconnects

Divya Mishra, Shailendra Mishra, Praggya Agnihotry and B.K.Kaushik

**Abstract-** Crosstalk in VLSI interconnects is a major constrain in DSM and UDSM technology. Among various strategies followed for its minimization, shield insertion between Aggressor and Victim is one of the prominent options. This paper analyzes the extent of crosstalk in inductively coupled interconnects and minimizes the same through distributed shield insertion. Comparison is drawn between signal voltage and crosstalk voltage in three different conditions i.e. prior to shield insertion, after shield insertion and after additional ground tap insertion at shield terminal.

——————————— ◆ ———————————

## I. INTRODUCTION

The rapid advances in VLSI (Very Large Scale Integration) technology has resulted in the reduction of minimum feature size to sub-quarter microns and switching time in tens of Pico seconds or even less. As a result, digital circuits today face the same problems which were once subjective to analog circuits only i.e. noise.

The device noise like shot noise, flicker noise and thermal noise are still subdued phenomenon with respect to the performance of digital circuits. However, external noise sources like crosstalk, substrate noise, power and ground bounce significantly degrade the performance and reliability of digital circuits.

Interconnect delay and crosstalk presently dominate the performance and signal integrity of deep sub micrometer VLSI circuits as far as on chip interconnects are considered. The root cause of this abnormality is that as feature sizes are decreased to deep sub micrometer dimensions, coupling capacitances significantly affect the circuit performance. This results due to decreased interconnect spacing and increased interconnect thickness.

Coupling noise has two deleterious effects on integrated circuits.

In case of a static signal, the noise transiently destroys the logical information stored on the static node resulting in an incorrect machine state stored within a latch, resulting in a functional failure.

When noise is accompanied with a switching event, the effect of noise is manifested as a change in the timing of the signal transition [1].

With the increment in circuit density the RLC delay becomes a factor that limits the useable clock frequencies. The effect is tried to compensate by using improved materials (Cu conductors and low-k dielectrics), make more accurate calculations and take into full account the transmission line behavior of the conductors or even a three-dimensional description [3]. However, these strategies have little or no contribution in improving.

## 2. PRINCIPLE OF SHIELDING

Shielding in high speed digital circuits is one of the effective and common ways to reduce crosstalk noise and signal delay uncertainty. Shield is a wire directly connected to Vdd or Gnd. One of the effective methods of shielding is placing ground or power lines at the sides of a victim signal line to reduce noise and delay uncertainty. This can be easily explained with the help of fig 1.

Different parameters which are considered with shield insertion include the effects of the shield width, length, separation between the shield and the signal, and the number of connections tieing the shield to ground on the crosstalk.

————————————————

- *Divya Mishra is with the Department of Applied Science, Vidya College of Engineering, Meerut, India.*
- *Shailendra Mishra is with the Department of Electronics & Communication Engg. Meerut Institute of Engineering and Technology, Meerut, India.*
- *Praggya Agnihotry is with the Department of Electronics& Communication Engg. Subharti Institute of Technology & Engineering, Meerut, India.*
- *B.K. Kaushik is with the Department of Electronics and Computer Engg., Indian Institute of Technology-Roorkee, Roorkee-247667, India.*

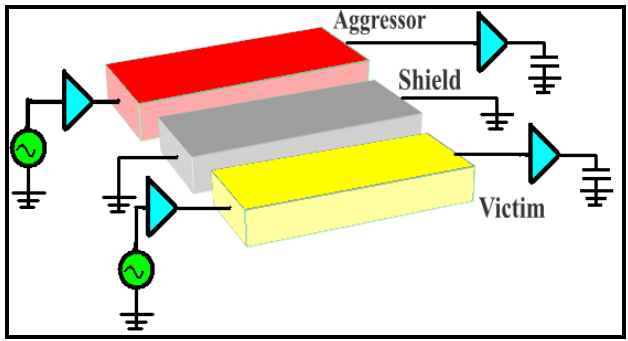

Fig. 1 Interconnect model with shield insertion

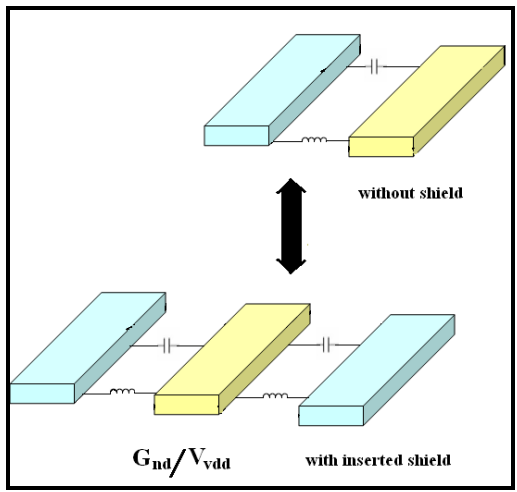

Fig.2. Circuit model of interconnect with inserted shield.

The above mentioned figure (Fig.2) shows that by inserting a shield line between the two adjacent signal lines, the coupling capacitance between the two signal lines disappear which is replaced by two new coupling capacitance between the signal line and shield line. Consequently, shield line efficiently isolates the voltage switching activities of the neighboring lines due to coupling capacitance. In addition, it also reduces inductive coupling noise. The shield line provides a closer and clearly defined current return path for both the signal lines [4], so the mutual inductive effect are reduced significantly compared to spacing strategy. But the picture is not as simple as it appears in case of inductive coupling as it can not be eliminated by just inserting a shield lines because the mutual inductance still exists between the two signal lines (Fig.2). The crosstalk noise for shielded interconnect increases as signal length increases and decreases with the increment in shield width [5].

Before the model is derived and analyzed there are few important facts which need to be mentioned about coupling capacitance and coupling inductance which exist in interconnect. These are the facts on the basis of which the model is developed.

**Capacitive Coupling:**

1. Coupling capacitance virtually exists only between adjacent wires or crossing wires and capacitive crosstalk is localized.

2. The crosstalk due to capacitive coupling affects both delay and signal integrity.

3. Capacitance can be pre-computed for a set of (localized) interconnect structures.

4. Coupling capacitance is highly sensitive to spacing.

5. Proper wire sizing and spacing may limit the impact of coupling capacitance by changing the ratio of coupling capacitance to total capacitance.

6. Reduction in coupling capacitance's impact may be initiated by increase the driver size of victim, decrease the driver size of aggressor, buffering, spacing, Net ordering and shielding.

**Inductive Coupling:**

1. On-chip inductance impacts have become more significant with the technology scaling and increase of clock frequencies.

2. On-chip inductance should be considered when ωL becomes comparable to R as we move towards Ghz+ designs.

3. Inductive effects can be minimized by staggered inverters/buffers differential signals i.e. Nets with opposite switching signals can be placed adjacent to each other, decrease inductive coupling noise at the cost of a higher capacitive coupling noise.

4. Inductive crosstalk is globalized

5. Inductive crosstalk affects both delay and signal integrity

6. Inductive crosstalk is not sensitive to spacing , wire sizing or Net ordering

7. Inductive crosstalk can be minimized by shielding, buffering, ground plane, differential signal or signal termination [4].

8. Inductive coupling exists between any two wires whereas capacitive coupling only exists between adjacent wires

It is observed that shielding is one of the efficient methods for both capacitive and inductive crosstalk reduction in VLSI interconnects. Inserting a shield line is necessary to retain proper signal integrity. However, shield insertion consumes more power, increases routing area and add to interconnect routing complexity [5].

## 3. PARAMETER CALCULATION

The model used for present simulation is a 2π RLC model where shield line has its two ends tied to ground rather than an ideal ground. Driving resistance is incorporated at the driver end and a load capacitance at the load end. The distributed parameters are gained through lumped circuit model. CMOS circuit is based on realistic assumptions and 90nm CMOS process technology is utilized. The model is based on symmetrical interconnect system oxide dielectric constant is 3.9 (SiO2), Load capacitance ($C_L$) is 76 fF, Transistor gate resistance is 82.76Ω Sheet resistance of metal for today's advance process technology is 30-50 mΩ/sq so the sheet resistance is assumed to be 50 mΩ/sq. Assuming the current only returns through the power and shield lines, the parameters related to the model are calculated as follows:

Rline = 500Ω

$$L_{line} = 0.002l\left[ In\left(\frac{2l}{w+t}\right) + 0.5 - In(\lambda) \right] \quad (1)$$

$$L_m = 0.002l\left[ In\left(\frac{l}{d} + \sqrt{\frac{l^2}{d^2}}\right) - \sqrt{1 + \frac{d^2}{l^2}} + \frac{d}{l} \right] \quad (2)$$

$$C_{line} = \varepsilon\left[ \left(\frac{w}{h}\right) + 0.77 + 1.06\left(\frac{w}{h}\right)^{0.25} + 1.06\left(\frac{t}{h}\right)^{0.5} \right] \quad (3)$$

$$C_m = \varepsilon_{ox}\left[ 1.035\left(\frac{w}{h}\right) + 1.83\left(\frac{t}{h}\right)^{-.22} - 1.07\left(\frac{t}{h}\right)^{-.22} \right]\left(\frac{d}{h}\right)^{.34} \quad (4)$$

Where,

l= Length of the interconnect = 5000 μm

t= Metal thickness. =2μm

w= width

d= Separation of track from nearest neighbors.

 = 2μm & 1μm on shield insertion w.r.t to shield
h=track substrate height = 2μm

εox = relative permittivity of SiO2 dielectric

= 3.9×εo = 3.9x8.86×10-12 Farad/meter

$C_{line}$=line capacitance

$C_m$=mutual capacitance

## 4. RESULTS

The simulation results are obtained for distributed RLC models without and with shield insertion as shown in figure 3 and 4 respectively. Both lines are capacitively and inductively coupled to each other.

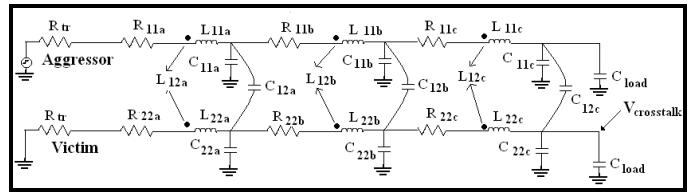

Fig3. Circuit model of interconnect with Aggressor and Victim.

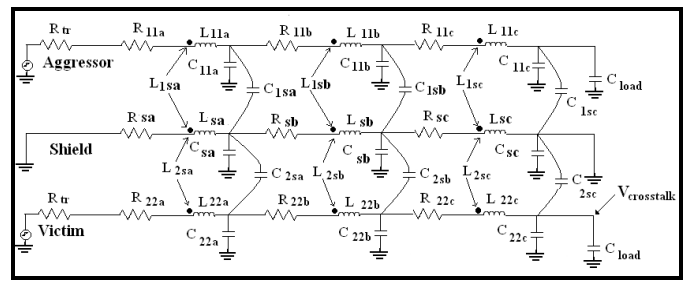

Fig4. Circuit model of interconnect with Aggressor and Victim .

$L_{line}$ (line inductance), $L_m$ (mutual inductance), $C_{line}$ (line capacitance), and $C_m$ (coupling capacitance), with and without shield insertion scenarios are obtained using equations (1, 2, 3 & 4). Obtained values are tabulated in table 1:

TABLE1:

PARAMETERS OBTAINED AND USED FOR SIMULATION

| Parameter | Without Shield | With Shield insertion |
|---|---|---|
| $L_{line}$ | 83.24μH | 83.24μH |
| $L_m$ | 8.21μH | 7.51μH |
| $C_{line}$ | 134.41pF | 134.41pF |
| $C_m$ | 69.50pF | 27.47pF |

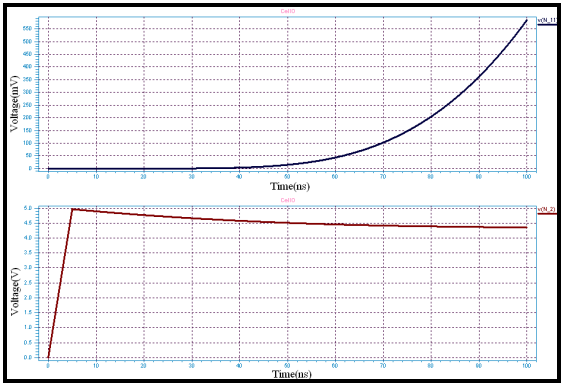

Fig. 5: Waveforms for Victim and Aggressor line with no shield insertion

The waveform in figure 5 and 6 shows clearly the extent of crosstalk voltage that exists on the victim line in absence and presence of the ground shield respectively. In case of both victim and aggressor the peak values of signal voltages are indicated. On insertion of the shield, considerable reduction in crosstalk voltage is observed on the victim line.

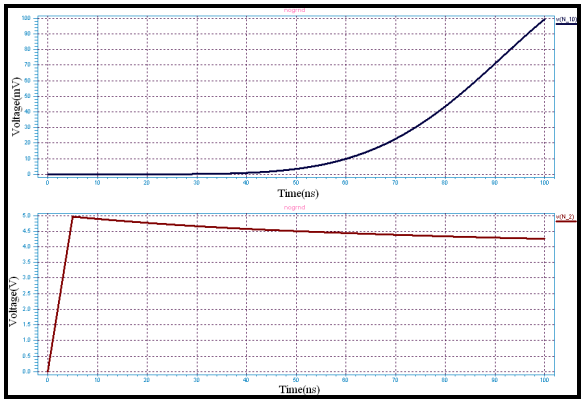

Fig.6: Waveforms for Victim and Aggressor model with shield insertion

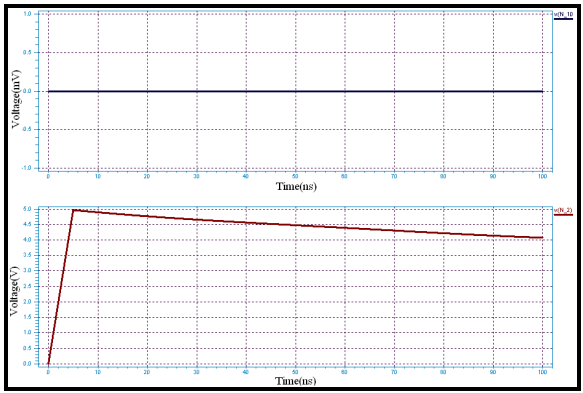

Fig7: Waveforms for Victim and Aggressor lines with three ground taps in shield

Now the analysis is extended further by addition of three taps to ground line on the shield line. This divides interconnect equally into three parts. It is observed that insertion of taps completely removes the crosstalk voltage on victim line. On the basis of the simulation results it is inferred that ground taps are extremely useful methodology for crosstalk elimination.

Table 2 depicts the effect of shield insertion on propagation delay and rise times on aggressor and victim lines. The delay and transition time on victim lines corresponds to noise signal. As crosstalk noise dies down due to shield insertion and ground taps the propagation delay and rise time on victim line also reduces. The delay and transition time on aggressor line also reduces with crosstalk reduction. Table 2 explicitly shows these delay and rise time changes in the victim signal and aggressor noise signal.

TABLE 2:

CROSSTALK VOLTAGE, DELAY AND RISE TIME VALUES WITH AND WITHOUT SHIELD

| $V_{aggressor}$ | | | |
|---|---|---|---|
| Parameter | No Shield | Shielded | Shield with 3 Gnd. Connections |
| $V_{victim}$ | 590mV | 100mV | 0fV |
| Delay $_{aggr.}$(ns) | 52.86 | 51.97 | 51.57 |
| Delay $_{vic.}$(ns) | 90.24 | 31.8 | 0 |
| Aggressor Rise time(ns) | 6.88 | 5.96 | 4.3 |
| Victim Rise time(ns) | 97.98 | 27.5 | 0 |

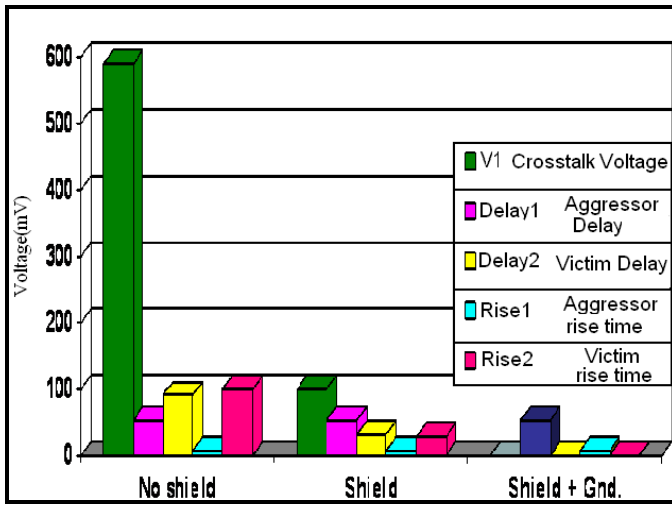

Fig 8: Crosstalk Voltage variations due to shields

Figure 8 shows the variations in crosstalk voltage in different situations of shields and without shield.

## 5. CONCLUSION

On basis of the distributed RLC model and simulations performed it is clear that the extent of crosstalk voltage decreases with the insertion of shield between the aggressor and victim line. The ground connection in a shield line divides the interconnect structure into smaller interconnect structures thereby, further reducing the crosstalk voltage. The simulation results shows that with the insertion of three ground tap connections in shield the effect of crosstalk i.e. crosstalk voltage on the victim line is completely eliminated.